%% file: final.tex
\documentclass[12pt]{article}
\usepackage{epsfig}
\usepackage{hhline}
\def\be {\begin{equation}}
\def\ee {\end{equation}}
\def\bea{\begin{eqnarray}}
\def\eea{\end{eqnarray}}
\def\x{\times}

\begin{document}
\renewcommand{\arraystretch}{1.5}

\rightline{UG-2/97}
\rightline{hep-th/9704120}
\rightline{April 1997}
\vspace{1.5truecm}
\centerline{\bf INTERSECTIONS INVOLVING MONOPOLES} 
\centerline{\bf AND WAVES IN ELEVEN DIMENSIONS}
\vspace{1.5truecm}
\centerline{\bf E.~Bergshoeff, \ M.~de  Roo, \  E.~Eyras,}
\centerline{\bf B.~Janssen \ and \ J.~P.~van der Schaar}
\vspace{.4truecm}
\centerline{{\it Institute for Theoretical Physics}}
\centerline{{\it Nijenborgh 4, 9747 AG Groningen}}
\centerline{{\it The Netherlands}}
\vspace{3truecm}
\centerline{ABSTRACT}
\vspace{.5truecm}

We consider intersections in eleven dimensions involving
 Kaluza-Klein monopoles and Brinkmann waves. Besides these
 purely gravitational configurations
 we also construct solutions to the equations of motion
 that involve  additional $M2$- and $M5$-branes.
 The maximal number of independent objects in these intersections
 is nine, and such maximal configurations, when reduced to two dimensions,
 give rise to a zero-brane solution with dilaton coupling $a=-4/9$.

\newpage

\noindent{\bf 1. Introduction}
\vspace{3mm}

Eleven dimensional supergravity has regained its prominent role
 in the search for a quantum theory of gravity. 
 It is the low-energy limit of the conjectured $M$-theory, from which
 all five ten-dimensional string theories can be obtained.
 
\begin{figure}[h]
\centering
\input{M-A.pstex_t}
\bigskip
\caption{{\scriptsize 
 {\bf The relation between $D=10$ IIA and $D=11$ solutions:}
 Vertical lines imply direct dimensional reduction, diagonal
 lines double dimensional reduction. The shadowed area indicates
 the relationship between known ten-dimensional solutions and a 
 conjectured 9-brane in $D=11$.}}
\label{M-IIA}
\end{figure}
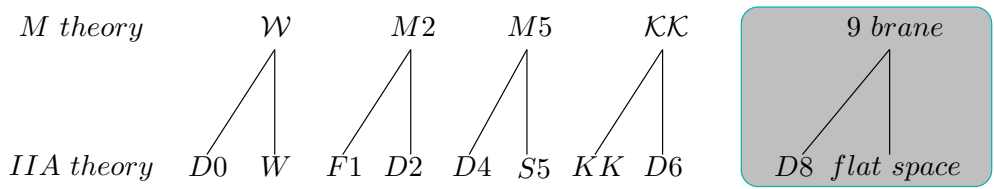

One implication of this viewpoint is, that all solutions of Type IIA 
 theory should have an eleven dimensional interpretation \cite{Towns}.
 Indeed, the fundamental string ($F1$) \cite{Dabh}
 and the solitonic five-brane
 ($S5$) \cite{Callan,Duff-Lu} are the double dimensional reduction of 
 the eleven dimensional $M2$-brane \cite{Duff-Stelle}
 and the direct dimensional reduction of the 
 eleven dimensional $M5$-brane \cite{Guven}, respectively.
 The Dirichlet $D2$- and $D4$-branes can be 
 obtained from $M2$ and 
 $M5$ via direct and double dimensional reduction, respectively.
 The $D0$- and $D6$-branes in the IIA theory are
 related to the purely gravitational 
 Brinkmann wave \cite{Brink} ($\cal W$) and 
 the Kaluza-Klein monopole \cite{Sorkin} ($\cal KK$) in eleven dimensions.
 These eleven dimensional solutions also have their counterparts
 in $D=10$, which we denote by $W$ and $KK$.
 Each of these solutions preserves 1/2 of the $D=11$ 
 (or $D=10\ N=2$) supersymmetry.
 In Figure \ref{M-IIA} we summarize the relationship between these
 $D=10$ IIA and $D=11$  solutions. 
 The eleven 
 dimensional interpretation of the Type IIA 8-brane 
 \cite{Pol2,Be2} is still a 
 mystery (see also below). Presumably, it is related  to a 9-brane
 \footnote{The conjectured 9-brane is also
 discussed in \cite{Be2,Howe,Papa2,Pol1,Duff}.}
 in $D=11$. The direct reduction of such a 9-brane
 is expected to lead to $D=10$ Minkowski space. 

\vspace{3mm}
\vspace{3mm}

The aim of this paper is to extend our recent work 
 on intersections of $M2$- and $M5$-branes \cite{mult} by including
 the wave and monopole solutions indicated in Figure 1.
 This paper
 is organized as follows. In Section 2 we will first discuss the case of
 two intersecting eleven dimensional solutions.
 In Section 3
 we obtain all multiple intersections which are purely gravitational,
 i.e., which do not involve the 3-form gauge field of $D=11$
 supergravity. In Section 4 we discuss multiple intersections involving
 $M2$- and $M5$-branes as well. 
 We draw our conclusions in Section 5.
 In the remainder of this Section we will summarize some relevant
 properties of the $\cal W$ and $\cal KK$ solutions.

The Brinkmann wave  in $D$ dimensions 
 is given by the metric
\be
\label{wave}
ds^2= (2-H)dt^2 -Hdz^2 + 2(1-H)dtdz - (dx_2^2 + ... +dx_{(D-1)}^2), 
\ee
where $H$ is a harmonic function in the variables $t+z, 
 x_2,\ldots, x_{(D-1)}$. In ten dimensions the wave solution is
 $T$-dual to the fundamental string $F1$\footnote{Since the
 wave, and also the monopole solution considered below, involve only
 fields which IIA and IIB theories have in common, this 
 duality transformation can be considered as a IIA transformation.},
 after assuming isometry in the $z$-direction.

There are two ways to reduce the wave to $D-1$
 dimensional spacetime. On imposing that $z$ is an isometry direction,
 the solution becomes static and corresponds in $D-1$ dimensions
 to a 0-brane.
 The charge is carried by a vector field of which only the time
 component does not vanish, and is given by $A_t= 1- H^{-1}$.
 Alternatively, one can impose that $H$ is independent of one of the
 $x_\mu\ \bigl ( \mu = 2,\cdots , (D-1)\bigr )$ coordinates. 
This results in a Brinkmann wave in $D-1$
 dimensions.

The metric for the Kaluza-Klein monopole reads ($i=1,2,3$)
\be
\label{monopole}
ds^2 = dt^2 - dx_1^2 - ... -dx_{(D-5)}^2 -H^{-1}(dz + A_i dy_i)^2 
      - H dy_i^2\,,  
\ee
where $H$ and $A_i$ depend on $y_i$, and 
the relation between $H$ and $A_i$ is
\be
F_{ij} \equiv \partial_i A_j - \partial_j A_i =  \epsilon_{ijk}\partial_k H\,.
\ee
Here the directions $t,x_\mu\ \bigl (\mu = 1,\cdots ,(D-5)\bigr)$ and 
$z$ are isometry directions. Reduction
 over $x_\mu$ leads to a Kaluza-Klein monopole in $D-1$ dimensions.
 Reduction over $z$ leads to a $(D-5)$-brane in $D-1$ dimensions, where
 the $y_i$-directions correspond to the transverse space. 
 The solution (\ref{monopole}) in ten dimensions is $T$-dual,
 with respect to the $z$-direction, to the
 solitonic fivebrane $S5$.

At several occasions we will assume
 that one of the $y_i$, say $y_1$, corresponds to an
 isometry direction as well. 
 In that case $A_2$ and $A_3$ can be gauged away, and 
 the metric becomes (in $D-1$ dimensions)
\be
  ds^2 = \varphi^{-1/2}(dt^2 - dx_1^2 - ... -dx_{(D-5)}^2)
     - \varphi^{1/2}(dz^2 + (H^2+A_1^2)(dy_2^2 + dy_3^2))\,,
\ee
where $H,\ A_1$ and 
\be
\varphi\equiv H/(H^2+A_1^2)
\ee
 are harmonic in $y_2,y_3$.
 The coordinate transformation to $u,v$, where
\be
\label{gct}
    d(u+iv) = (H+iA_1)\,d(y_2 +iy_3)
\ee
preserves the harmonic property of $\varphi$, and gives the usual 
 metric, dilaton and a vector field with a non-vanishing component
 in the $z$-direction for a magnetic $(D-5)$-brane. 
 Of course $z$ remains an isometry direction in
 $D-1$ dimensions. The coordinate transformation (\ref{gct}) can
 also be done directly in $D$ dimensions. 

Sometimes we will consider monopoles which are 
 truncated further, and for which the harmonic function $H$
depends on only a
 single variable, say $y_3$. This implies that (locally) $H=my_3+c$,
 $A_1=-my_2$ for constant $c$. 
Note that this does not imply an additional isometry,
 and reduction over $y_1$ indeed gives, after the coordinate transformation
 (\ref{gct}), a $(D-5)$-brane for which $\varphi$ again 
 depends on $u$ and $v$.

To obtain a $(D-5)$-brane in $D-1$ which has two additional
 isometries we must choose for $H$ and $A_1$ special functions that
are harmonic in $y_2, y_3$.
 If the harmonic function $\varphi$ depends only on $u$, it must be
 linear in $u$, and the coordinate transformation (\ref{gct}) then
 implies that $H$ and $A_1$ must satisfy
\be
    d\,\left( {H-iA_1\over H^2+A_1^2}\right) = (H+iA_1)d(y_2+iy_3)\,,
\ee
which is solved by
\be
    (H+iA_1)^2 = {1\over 2(y_2+iy_3+\alpha)}\,,
\ee
where $\alpha$ is a complex integration constant.
 Reducing over the $y_1$ direction, we find that indeed
 the function $\varphi$, after the coordinate transformation (\ref{gct}),
 depends on $u$ only, and that the
 only nonzero component of the gauge field 
 is in the $z$-direction, and is given by $v d\varphi/du$.
 This form of the $(D-5)$-brane 
 in $D-1$ dimensions was given in \cite{Towns} 
 for the case of the 6-brane in ten dimensions. There it is $T$-dual
 to the 8-brane \cite{Be2}. Note that strictly speaking the $(D-5)$-brane
does not have two additional isometries since the gauge field is linear in $v$.
However, as discussed in \cite{Be2}, such linear dependence disappears
after a further reduction over $v$ to $D-2$ dimensions. Furthermore, the
$v$--dependence also disappears in the $D-1$ dimensional dual formulation 
where the
vector field has been replaced by a $(D-4)$--form gauge field.
In this sense we may consider the 
$v$-direction as a kind of ``generalized'' isometry direction.

It is interesting to consider the
uplifting of the truncated $(D-5)$-brane solution discussed above
to $D$ dimensions:

\be 
ds^2 = dt^2 - dx^2_1 - \dots - dx^2_{(D-5)} - u^{-1}\bigl (
dy -  vdz \bigr )^2 - u \bigl ( dz^2 + du^2 + dv^2 \bigr )\, .
\ee
Since this solution has $D-2$ isometries and one ``generalized''
isometry, it is similar to a $(D-2)$-brane solution in $D$ dimensions.
For $D=11$ this would correspond to a 9-brane solution. Upon reduction to
8 dimensions it leads to a solution which is identical to the
ten-dimensional 8-brane when reduced to 8 dimensions.

The Kaluza-Klein monopoles, for which additional isometry is imposed in the
 direction of the Kaluza-Klein vectors, are no longer asymptotically flat.
 Although this will disqualify them for certain applications, they are 
 nevertheless solutions of the equations of motion, and reduce to
 (truncated) $D6$-branes in $D=10$. Since in this paper we do not consider
 global properties of our solutions, we will include 
 these truncated monopoles in multiple intersections.

\vspace{3mm}
\noindent{\bf 2. Intersection rules}
\vspace{3mm}

Intersections of a pair of branes are at the basis of the construction
 of multiple intersections. In a multiple
 intersection each pair  obtained by setting all but two of the
 independent harmonic functions equal to one, must be one of the basic 
 pairs described below. For the $Dp$-branes and the NS-NS-branes
 $F1$ and $S5$ in $D=10$, as well as for the $M2$ and $M5$ branes
 in $D=11$, the allowed pair intersections are known
 \cite{Papad-Towns1,Tseyt1, BeBeJa, Gaunt, Tseyt2, Are}. For pair intersections
 involving waves and monopoles partial results were given before
 \cite{Tseyt1,Khvien,Costa}. 

In Tables 1 and 2 we summarize old and new results on the pair intersections.
 The two independent harmonic functions of the pairs in Table 1
 depend on the coordinates which are transverse to both branes
 (overall transverse)\footnote{For some of the entries in Table 1
 another possibility exists, namely that
 one harmonic function depends on overall transverse, 
 the other on directions
 which are transverse to only one brane in the pair (relative transverse) 
 \cite{BeBeJa, Tse}.
 We will not consider this option in this paper.}.
 For the pairs in Table 2 both harmonic
 functions must depend on the relative transverse coordinates.
 In Sections 3 and 4, where we discuss
 multiple intersections,  we will use only
 the pairs of Table 1.

The first three rows of Table 1 denote the intersections of $M2$- and
 $M5$-branes. As an example, which also explains our notation, consider
 $(1|M2,M5)$. Denoting a worldvolume direction of a brane by $\x$, and
 a transverse direction by $-$, the metric for this pair can be
 represented by 

\bea
 (1|M2,M5) &=& 
\mbox{ {\scriptsize
        $\left\{ \begin{array}{c|cccccccccc}
         \x & \x & \x &  - &  - &  - &  - & - & -  & -  & - \\
         \x & \x &  - & \x & \x & \x & \x & - & -  & -  & -             
                         \end{array} \right.$ } }   \label{M2M5} 
\eea
The coordinates $t=x^0,x^1,\ldots x^{10}$ are indicated from 
 left to right. The common worldvolume in this case is two-dimensional
 ($x^0,x^1$),
 the overall transverse space four-dimensional ($x^7,\ldots,x^{10}$), 
 and there are five relative transverse 
 directions ($x^2,\ldots,x^6$). 
 The space-like directions
 $x^1,\ldots,x^6$ correspond to isometries. Reduction over $x^1$
 gives $(0|F1,D4)$ in ten dimensions. For the relative transverse
 directions the possibilities are: either reduction over $x^2$,
 giving $(1|F1,S5)$, or reduction over one of the directions
 $x^3,\ldots,x^6$, giving $(1|D2,D4)$. Finally, one can impose
 an isometry in one of the overall transverse directions by restricting
 the dependence of the harmonic functions to three coordinates. Reduction
 over such a direction gives $(1|D2,S5)$.
The next two rows represent the addition of a wave to the
 $D=11$ $M$-branes. The $z$-direction of the wave must be placed in the
 world volume of the $M$-brane. The dependence of the harmonic functions
 is only on the directions transverse to the $M$-brane, so that
 the wave does not propagate.
 The metric for these two $D=11$ pairs can be
 represented by\footnote{Note that we extend the
 notation $(q|p_1,p_2)$ to include waves and monopoles with the
 understanding that the worldvolume directions
 of the ``${\cal W}$-brane'' are given by $t,z$ (see (\ref{wave})),
 and the
 transverse directions of the ``${\cal KK}$-brane'' are given by
 the isometry direction $z$ and the coordinates in which the
 Kaluza-Klein vector is oriented. These directions (called $y_i$ in
 (\ref{monopole})) will be denoted by $A_i$.}

\bea
 (1|M2, {\cal W})&=&
\mbox{ {\scriptsize
$\left\{ \begin{array}{c|cccccccccc}
         \x & \x & \x & -  & -  & -  & -  & - & -  & -  & - \\           
         \x &  z & -  & -  & -  & -  &  - & - & -  & -  & - 
                                         \end{array} \right.$}} \label{WM2}  \\
 (1|M5, {\cal W})&=&
\mbox{ {\scriptsize
$\left\{ \begin{array}{c|cccccccccc}
         \x & \x & \x & \x & \x & \x & -  & - & -  & -  & - \\           
         \x &  z & -  & -  & -  & -  &  - & - & -  & -  & - 
                                         \end{array} \right.$}}  \label{WM5}  
\eea

\begin{eqnarray*}
\begin{array}{|c||c|c|c|}
\hline
\mbox{}&\mbox{common wv.}&\mbox{relative trv.}&\mbox{overall trv.}\\
\hline
\hline
  (0|M2,M2)    &      -          & (0|F1,D2)          & (0|D2,D2)    \\
\hline
  (1|M2,M5)    &(0|F1,D4)        & (1|F1,S5)          & (1|D2,S5)     \\
               &                 & (1|D2,D4)          &               \\       
\hline 
  (3|M5,M5)    &(2|D4,D4)        & (3|D4,S5)          & (3|S5,S5)     \\ 
\hline 
\hline
(1|M2,{\cal W})&(0|F1,D0)        & (1|F1,W)           & (1|D2,W)     \\      
\hline 
(1|M5,{\cal W})&(0|D4,D0)        & (1|D4,W)           & (1|S5,W)     \\      
\hline
\hline
(2|M2,{\cal KK})&(1|F1,KK)       & (2|D2,KK)          & (2|D2,D6)     \\ 
\hline 
(5|M5,{\cal KK})&(4|D4,KK)       & (5|S5,KK)          & (5|S5,D6)    \\ 
\hline
(0|M2,{\cal KK})& -              & (0|F1,D6)          & (0|D2,D6)^*  \\
                &                & (0|D2,KK)          &              \\
\hline
(3|M5,{\cal KK})&(2|D4,KK)       & (3|D4,D6)          & (3|S5,D6)^*   \\ 
               &                 & (3|S5,KK)          &               \\
\hline
\hline 
(1|{\cal W},{\cal KK})&(0|D0,KK) & (1|W,KK)           & (1|W,D6)     \\      
\hline
(4|{\cal KK},{\cal KK})^a&(3|KK,KK)^a& (4|D6,KK)^*    & (4|D6,D6)     \\ 
\hline  
(4|{\cal KK},{\cal KK})^b&(3|KK,KK)^b&  (4|D6,KK)     & (4|D6,D6)^*    \\ 
\hline 
\end{array}\nonumber
\end{eqnarray*}

\vspace{3mm}
{\scriptsize
\noindent Table 1.\ {\bf Pair intersections in $D=11$ and their
 reductions to $D=10$
 with dependence on overall transverse coordinates:} 
 The first column represents the
 pair intersections in $D=11$. $(q|p_1,p_2)$
 denotes an intersection of a $p_1$ and a $p_2$ brane 
 over a common $q+1$-dimensional worldvolume. Reductions to nontrivial
 solutions in $D=10$,
 obtained by compactification in different directions 
 (common worldvolume, relative transverse and overall transverse) with
 respect to the branes, are
 indicated in the remaining columns.  The $D=10$ solutions marked with ${}^*$
 are not of the usual harmonic form.
}
\vspace{5mm}

The next four rows in Table 1 denote the pairs involving
 one $M$-brane and one Kaluza-Klein monopole. 
 The metric for these four cases takes on the form
\bea
 (2|M2, \cal{KK})&=&
\mbox{ {\scriptsize
   $\left\{ \begin{array}{c|cccccccccc}
         \x & -  & -  &  - & -  & \x & \x & - & -  & -  & - \\
         \x & A_1& A_2& A_3& z  & \x & \x & \x & \x & \x & \x 
                                   \end{array} \right.$}}   \label{M2KKtr} \\
 (5|M5,\cal{KK})&=&
\mbox{ {\scriptsize
$\left\{ \begin{array}{c|cccccccccc}
         \x & -  & -  & -  & -  & \x & \x & \x &\x  & \x & - \\
         \x & A_1& A_2& A_3&  z & \x & \x & \x & \x & \x & \x  
                                    \end{array} \right.$}} \label{M5KKtr}\\
 (0|M2, \cal{KK})&=&
\mbox{ {\scriptsize
$\left\{ \begin{array}{c|cccccccccc}
         \x & \x & -  &  - & \x & - &  - & - & -  & -  & - \\
         \x & A_1& A_2& A_3&  z &\x & \x & \x& \x & \x & \x            
                                    \end{array} \right.$}}   \label{M2KKwv} \\
 (3|M5, \cal{KK})&=&
\mbox{ {\scriptsize
$\left\{ \begin{array}{c|cccccccccc}
         \x & \x &  - & -  & \x & \x & \x & \x & -  & -  & - \\
         \x & A_1& A_2& A_3& z  & \x & \x & \x & \x & \x & \x 
                                    \end{array} \right.$}} \label{M5KKwv} 
\eea
As we see, there are two 
 possibilities. The $z$-direction of the
 Kaluza-Klein monopole, the natural isometry direction
 which on compactification gives a magnetic $(D-5)$-brane, can be
 placed either in a direction transverse to 
 ($(2|M2,{\cal KK})$ and $(5|M5,{\cal KK})$) or in the
 worldvolume of the $M$-brane ($(0|M2,{\cal KK})$ 
 and $(3|M5,{\cal KK})$).
 The solutions (\ref{M2KKtr}) and (\ref{M5KKtr}) have been given before in 
 \cite{Tseyt1, Costa}. For these, the reduction to $D=10$ is straightforward.
 Note that the reduction over an overall transverse direction
 can be either over a direction indicated by $z$, or, by imposing an 
 additional isometry, in the direction of a component of the vector
 field.

In the solutions (\ref{M2KKwv}) and (\ref{M5KKwv}) 
 the harmonic functions
 depend only on the two overall transverse coordinates, 
 so that the Kaluza-Klein monopole
 has one additional isometry direction (indicated by $A_1$).
 In both of these solutions the reduction over the relative
 transverse $A_1$ and $z$
 directions yields, after a coordinate transformation, the same result.
 
The last three rows of Table 1 correspond to
 intersections of Kaluza-Klein monopoles and
 waves. The possibilities are shown in (\ref{WKK}-\ref{KKKKb})\footnote{
 Solution (\ref{WKK}) was presented in \cite{Tseyt1}.}.
 Note that there are two ways to intersect two Kaluza-Klein 
 monopoles, both with a five-dimensional common worldvolume. In solution
 (\ref{KKKKa}) the two harmonic functions depend on a single coordinate
 ($x^1$),
 in (\ref{KKKKb}) on two coordinates ($x^1,x^{2}$).
 
\bea
(1|{\cal W}, {\cal KK})&=&
\mbox{ {\scriptsize
$\left\{ \begin{array}{c|cccccccccc}
         \x &  - & -  & -  & -  &z_1  &  - & - & -  & -  & - \\
         \x & A_1& A_2& A_3& z_2&\x & \x & \x & \x & \x & \x 
                                         \end{array} \right.$}} \label{WKK}\\
 (4|{\cal KK},{\cal KK})^a &=& 
\mbox{ {\scriptsize
$\left\{ \begin{array}{c|cccccccccc}
         \x & A_1& A_2& A_3& z  & \x & \x & \x & \x & \x & \x  \\ 
         \x & B_1& \x & \x & z  & B_5& B_6& \x & \x & \x & \x     
 \end{array} \right.$}} \label{KKKKa}\\
 (4|{\cal KK},{\cal KK})^b &=& 
\mbox{ {\scriptsize
$\left\{ \begin{array}{c|cccccccccc}
         \x & A_1& A_2& A_3& z_1  & \x & \x & \x & \x & \x & \x  \\
         \x & B_1& B_2& \x & \x & B_5& z_2 & \x& \x & \x& \x  
                                         \end{array} \right.$}} \label{KKKKb}
\eea

For these solutions it may be useful to present the metric explicitly.
 We have:
\bea
  && (4|{\cal KK},{\cal KK})^a \to
   ds^2 = dt^2 - H_1H_2 dx_1^2 - H_1dx_{(2-3)}^2 - H_2dx_{(5-6)}^2
               - dx^2_{(7-10)} \nonumber \\
\label{KKa}        &&\qquad   
 - (H_1H_2)^{-1}(dz + (A_1+B_1)dx_1+A_2dx_2 
               + A_3dx_3+ B_5dx_5+B_6dx_6 )^2 \,, \\
  && \nonumber\\
  && (4|{\cal KK},{\cal KK})^b \to
   ds^2 = dt^2 - H_1H_2dx^2_{(1-2)} - H_1dx_3 - H_2dx_5  - dx^2_{(7-10)} 
 \nonumber\\
  &&\qquad\qquad\qquad
   - H_1^{-1}(dz_1 + A_1dx_1 + A_2 dx_2 + A_3dx_3)^2  \nonumber\\
\label{KKb}  &&\qquad\qquad\qquad\qquad
     - H_2^{-1}(dz_2 + B_1dx_1 + B_2 dx_2 + B_5dx_5)^2
  \,.
\eea
Note that in (\ref{KKa}) the harmonic functions depend only on $x_1$.
 Therefore 
 two of the components of each of the gauge fields $A$ and $B$ 
 can be gauged to zero. For the  
 reductions in Table 1 different gauge choices are employed. In
 (\ref{KKb}) the harmonic functions depend on $x_1$ and $x_2$. Here also
 different gauge choices can be made.

The solution (\ref{KKKKa}) solves the equations
 of motion, since it is the known ten-dimensional solution
 $(4|D6,D6)$ lifted up to $D=11$. The configuration (\ref{KKKKb}) 
 must be a solution because, after reduction over a common worldvolume
 direction it can be related to a known solution involving two
 solitonic five-branes via the following $T$-duality
 chain in $D=10$:
\be
(3|S5,S5) \rightarrow (3|S5,KK) \rightarrow (3|KK,KK)^b\,.
\ee
Note that it is possible to relate (\ref{KKKKa}) and (\ref{KKKKb})
 by a chain of $T$-duality and one $S$-duality transformation
 in ten dimensions. This involves the $S$-duality transformation
 between $(3|D5,D5)$ and $(3|S5,S5)$. 

Similarly, the intersection of a wave and a Kaluza-Klein monopole
 can be obtained from ten dimensions by first constructing
 an intersection in $D=10$ of a $D0$-brane with the Kaluza-Klein
 monopole:
\be
(0|D1,S5) \rightarrow (0|D0,KK)\,,  
\ee  
and by lifting this to eleven dimensions.

In Table 1 there are four reductions to $D=10$ that do not lead to
 solutions which are expressed in a standard form in terms of
 harmonic functions. As an example, consider the reduction of
 (\ref{KKKKa}). The harmonic 
 functions depend on $x_1$, the nonzero gauge field components can be 
 chosen to be $A_2$ and $B_5$, which then
 depend on $x_3$ and $x_6$, respectively. Reduction over $z$ gives 
 $(4|D6,D6)$, but also reduction over $x_2$ is possible. This gives
 a $D=10$ configuration which has the properties of $(4|D6,KK)$,
 but the fields do not have the standard harmonic form. It is
 given by:
\bea
    && ds^2 = \varphi^{-1/2}(dt^2 - dx^2_{(7-10)} -H_2dx^2_{(5-6)}) \nonumber\\
    &&\quad   -H_2^{-1}\varphi^{1/2}\left((dz+B_5dx_5)^2
          + (H_1^2H_2+A_2^2)(dx^2_3+H_2dx^2_1)\right)\,, \\
    && e^{2\phi} = \varphi^{-3/2}\,,\nonumber\\
    && C_z = {\varphi A_2\over H_1H_2}\,,\qquad
       C_5 = {\varphi A_2B_5\over H_1H_2}\,,\nonumber
\eea
 where
\be
  \varphi = H_1H_2/(A_2^2+H_1^2H_2)\,.
\ee
 The nonzero components of the RR-vector field in $D=10$ are denoted
by $C_\mu$.
 Note that $\varphi$ is indeed not harmonic in $x_1,x_3$. 
 If $H_2=1$ and $\ B_5=0$,  $\varphi$ does
 become harmonic, and we obtain a standard $D6$ solution,
 after the coordinate transformation (\ref{gct}). Conversely, for
 $H_1=1,\ A_2=0$ a standard Kaluza-Klein monopole is obtained in $D=10$.
 These solutions show that the usual harmonic Ansatz for intersecting pairs
 does not cover all possibilities. It will be interesting to
 investigate these non-harmonic solutions further (see also \cite{Tse}).  

\begin{eqnarray*}
\begin{array}{|c||c|c|c|}
\hline
\mbox{}&\mbox{common wv.}&\mbox{relative trv.}&\mbox{overall trv.}\\
\hline
\hline
(1|M5,M5)               &  (0|D4,D4)& (1|D4,S5) & (1|S5,S5) \\
\hline
\hline
(0|M2,{\cal {KK}})      &     -     & (0|D2,KK) & (0|D2,D6) \\
                        &           & (0|F1,D6)^* &         \\
\hline
(1|M5,{\cal KK})        & (0|D4,KK) & (1|S5,KK) &     -     \\
                        &           & (1|D4,D6) &           \\
\hline
(3|M5,{\cal KK})        & (2|D4,KK) & (3|S5,KK) &  (3|S5,D6) \\
                        &           & (3|D4,D6)^* &           \\
\hline
\hline
(2|{\cal KK},{\cal KK}) & (1|KK,KK) & (2|D6,KK) &     -     \\
\hline
(4|{\cal KK},{\cal KK}) & (4|KK,KK) & (4|D6,KK) &  (4|D6,D6)^*   \\
                        &           & (4|D6,KK)^* &         \\
\hline
\end{array}\nonumber
\end{eqnarray*}

\vspace{3mm}
{\scriptsize
\noindent Table 2.\ {\bf Pair intersections in $D=11$ and their
 reductions to $D=10$
 with dependence on relative transverse coordinates.} 
 The reductions indicated by a ${}^*$ are not expressed in a standard way
 in terms of harmonic functions.
}
\vspace{5mm}

In Table 2 we  consider intersections in which the two harmonic functions 
 depend on the relative coordinates. 
 There is one pair involving only $M5$ \cite{Gaunt}, and five
 pairs involving Kaluza-Klein monopoles. Some of these configurations
 and their generalization to non-orthogonal intersections
 were discussed recently in \cite{Tnew}.

Below we present the metric of these pairs in the usual way.
The pairs involving Kaluza-Klein monopoles are each related
 to known solutions through $D=10$, so that we can be sure that they solve
 the equations of motion.
For example,
 $(2|{\cal KK},{\cal KK})$ can be reduced to $(1|KK,KK)$ in ten dimensions and
 applying $T$-duality twice, in the directions $z_1$
 and $z_2$, we find
\begin{equation}
(1|KK,KK) \rightarrow (1|S5,KK) \rightarrow (1|S5,S5) \, ,
\end{equation}
and this can be oxidized to $(1|M5,M5)$, which is a known solution.

\begin{eqnarray}
  (1|M5, M5) &=&
\mbox{ {\scriptsize
       $\left\{  \begin{array}{c|cccccccccc}
                \x & \x & \x & \x & \x & \x & - & - & - & - & - \\
                \x & \x & - & -&  - & - &\x &\x &\x &\x & -
                        \end{array} \right.$}} \label{8M5M5}   \\
  (0|M2, \cal{KK}) &=&
\mbox{ {\scriptsize
        $\left\{  \begin{array}{c|cccccccccc}
                \x & \x & \x & - & - & - & - & - & - & - & - \\
                \x &A_1 &A_2 &A_3& z &\x &\x &\x &\x &\x &\x
                        \end{array} \right.$}} \label{8M2KK}   \\
  (1|M5,{\cal KK}) &=&
\mbox{ {\scriptsize
        $\left\{ \begin{array}{c|cccccccccc}
         \x &\x & \x & \x &\x  &\x & - & - & - & - & - \\
         \x &A_1 &A_2 &A_3 & z &\x &\x &\x &\x &\x & \x 
                        \end{array} \right.$}} \label{8M5KK} \\
  (3|M5,{\cal KK}) &=&
\mbox{ {\scriptsize
        $\left\{ \begin{array}{c|cccccccccc}
         \x &\x & \x & -   & - &-  & - & - &\x &\x & \x \\
         \x &A_1 &A_2 &A_3 & z &\x &\x &\x &\x &\x & \x 
                        \end{array} \right.$}} \label{83M5KK} \\
  (2|{\cal KK},{\cal KK}) &=&
\mbox{ {\scriptsize
        $\left\{ \begin{array}{c|cccccccccc}
         \x & A_1& A_2&A_3 & z_1& \x & \x &\x  & \x &\x &\x\\
         \x & \x & \x & \x & \x &z_2 & B_6& B_7& B_8& \x&\x 
                        \end{array} \right.$}} \label{8KKKK} \\
  (4|{\cal KK},{\cal KK}) &=&
\mbox{ {\scriptsize
        $\left\{ \begin{array}{c|cccccccccc}
         \x & A_1& A_2&A_3 & z_1& \x & \x  &\x  & \x &\x &\x\\
         \x & \x & \x &B_3 & B_4 &B_5 & z_2&\x  & \x &\x &\x 
                        \end{array} \right.$}} \label{84KKKK} 
\end{eqnarray}

In (\ref{8M5M5}-\ref{84KKKK}) the dependence is on the relative 
 transverse coordinates, e.g., in (\ref{83M5KK}) $H_1$
 depends on $x_5,\ldots,x_7$ and $H_2$ on $x_1,x_2$. In the reduction of
 (\ref{83M5KK}) to $D=10$ we obtain $(3|S5,KK)$ when an isometry in
 one of the coordinates $x_5,\ldots,x_7$ is assumed, and $(3|D4,D6)^*$
 when reducing over $x_1$ or $x_2$.

\vspace{3mm}
\noindent{\bf 3. Purely gravitational solutions: monopoles and waves}
\vspace{3mm}

In this Section we will consider configurations involving several
 monopoles, with or without an additional wave, using the
 pair intersections of Table 1. The interest of
 such solutions lies in the fact that they involve only the gravitational
 field. If the spacetime is of sufficient dimensionality, such
 solutions can always be present.

Configurations involving only monopoles differ in the way  the $z$-isometry
 directions are related. In (\ref{typeA}-\ref{typeC}) we present
 three configurations to which no further monopole  can be added.

\bea
&\mbox{Type A:}&
\mbox{ {\scriptsize
$\left\{ \begin{array}{c|cccccccccc}
         \x & \x   & \x   & \x & \x & \x & \x & A_7& A_8 & A_9  & z \\
         \x & \x   & \x   & \x & \x & B_5& B_6& \x & \x  & B_9  & z \\
         \x & \x   & \x   & C_3& C_4& \x & \x & \x & \x  & C_9  & z \\
         \x &D_   1&D_2   & \x & \x & \x & \x & \x & \x  & D_9& z 
                                         \end{array} \right.$}} \label{typeA}\\
&\mbox {Type B:}&
\mbox{ {\scriptsize
$\left\{ \begin{array}{c|cccccccccc}
         \x & \x & \x & \x & \x & \x   & \x   & A_7  & A_8  & A_9   & z_1 \\
         \x & \x & \x & \x & \x & B_5  &  B_6 & \x   & \x   & B_9   & z_2 \\
         \x & \x & \x & \x &  z_3 & C_5  & \x   & \x   & C_8  & C_9   &\x \\
         \x & \x & \x & \x &  z_4 & \x   & D_6  & D_7  & \x   & D_9   &\x \\
         \x & \x & \x &  z_5 & \x & E_5  &  \x  & E_7  & \x   & E_9   &\x \\
         \x & \x & \x &  z_6 & \x & \x   & F_6  & \x   & F_8  & F_9   &\x
                                         \end{array} \right.$}}\label{typeB} \\
&\mbox {Type C:}&
\mbox{ {\scriptsize
$\left\{ \begin{array}{c|cccccccccc}
         \x & z  & A_2& \x & \x & \x & \x  & \x & \x   & A_9  &A_{10} \\
         \x & \x & \x &  z & B_4& \x & \x  & \x & \x   & B_9  &B_{10} \\
         \x & \x & \x & \x & \x &  z & C_6 & \x & \x   & C_9  &C_{10} \\
         \x & \x & \x & \x & \x & \x & \x  &  z & D_8  & D_9  &D_{10}
                                        \end{array} \right.$}} \label{typeC}
\eea 

 In (\ref{typeA}) there is a common isometry direction $z$, in
 (\ref{typeB}) the six monopoles come in pairs with a common $z$-isometry,
 while the four monopoles in (\ref{typeC}) have no common $z$-isometry.
 Note that in (\ref{typeA}) and (\ref{typeB}) the solution depends
 on only one coordinate, in (\ref{typeC}) the harmonic functions may depend 
 on two coordinates. To the solution (\ref{typeB}) we can add a 
 single wave 
 in either the $x^1$ or the $x^2$ direction.

It is interesting to see how these purely gravitational solutions survive 
 in lower dimensions. In Table 3 we indicate the configurations
 with a maximum number of monopoles. Note that if we go to dimensions
 higher than 11, configurations of type A and type C are naturally
 extended to an additional monopole in each odd dimensional spacetime.
 The configurations of type B cannot be extended beyond six
 monopoles in higher dimensions.
 In some cases a single wave can be 
 added to these monopole configurations. Note that the solution in $D=5,6$ 
 is the same for type A, B, and C. In $D=7$ there is no difference 
 between type A and type B.

\begin{eqnarray*}
\begin{array}{|c||c|c|c|}
\hline
\mbox{D}&\mbox{type A}&\mbox{type B}&\mbox{type C}\\
\hline
\hline
  5    &     1         &    1       &   1     \\
\hline 
  6    &     1 + W     &    1 + W   &   1 + W    \\ 
\hline
  7    &     2         &    2       &   2     \\
\hline 
  8    &     2 + W     &    4       &   2 + W    \\
\hline
  9    &     3         &    6       &   3     \\
\hline 
  10   &     3 + W     &    6 + W   &   3 + W    \\
\hline
  11   &     4         &    6 + W   &   4     \\
\hline
\end{array}\nonumber
\end{eqnarray*}

\vspace{3mm}
{\scriptsize%
\noindent Table 3.\ {\bf Maximal number of monopoles and
 waves in $5\le D\le 11$ dimensions:} We indicate the maximum
 number of Kaluza-Klein monopoles in different dimensions,
 superimposed according to type A, B, or C (see
 (\ref{typeA}-\ref{typeC})). W means that a wave can be added. 
}
\vspace{5mm}

The supersymmetry of these purely gravitational solutions, 
 embedded in $D=11$
 supergravity and its toroidal compactifications, is 1/16 of the
 $D=11$ supersymmetry.
 
\vspace{3mm}
\noindent{\bf 4. Multiple intersections}
\vspace{3mm}
 
Having determined the ``no-force'' condition between the basic
 eleven dimensional solutions in Section 2 and the multiple
 intersections of waves and monopoles in Section 3, we next 
 consider multiple intersections that also involve $M2$- and $M5$-branes. 
 Multiple intersections of $D$-branes in $D=10$, and 
 of $M2$- and $M5$-branes in $D=11$ have only 
 recently been classified \cite{mult}. The $D=11$ result is
 given in Table 1 of \cite{mult}. In this Section we will generalize
 the result of \cite{mult} to intersections that 
 also involve waves and
 monopoles. 
 We will first restrict ourselves to configurations
 that can be reduced to intersections with only $D$-branes in $D=10$.
 Looking back at Table 1, we see that all pairs involving monopoles
 should then be of the form $(2|M2,{\cal KK})$, $(3|M5,{\cal KK})$
 or $(4|{\cal KK},{\cal KK})^a$, and that with a wave only
 $(1|M5,{\cal W})$ may be used.
 Thus  only multiple monopoles of Type A
 (see the previous Section) will be used. At the end of this Section we will
 relax these restrictions and consider the possibility of also
 using $(1|M2,{\cal W})$. 

Our strategy will be to take Table 1 of \cite{mult} as our starting
 point and then consider to which $M$-brane intersections waves and/or
 monopoles can be added. The rule for adding a wave is known
 \cite{Tseyt1,Tseyt3}. To each intersection involving at least a common string
 a wave can be added in such a way that the $z$-isometry direction
 of the wave lies in the spacelike common string direction.
 Furthermore, at most one wave can be added to any given intersection.

{}From the intersection (\ref{M2KKtr}) we see that the worldvolume of the
 $M2$-brane must lie in the worldvolume directions of the monopole.
 Furthermore two intersecting $M2$-branes have distinct (spacelike)
 worldvolume directions. Since the monopole has six (spacelike)
 worldvolume directions we conclude that monopoles may be added 
 to configurations that contain at most three $M2$-branes \cite{Costa}:

\be
\mbox{ {\scriptsize
$\left\{ \begin{array}{c|cccccccccc}
         \x & \x & \x &  - &  - &  - &  - & - & -  & -  & - \\
         \x &  - &  - &  \x&  \x&  - &  - & - & -  & -  & - \\
         \x &  - &  - &  - &  - &  \x& \x & - & -  & -  & - \\
         \x & \x & \x & \x & \x & \x & \x & z & A_8& A_9& A_{10}            
                                         \end{array} \right.$} } 
\ee

We next consider the $M5$-branes.
Using only the  pair $(3|M5,{\cal KK})$
 we see that the $z$-isometry direction
 of the monopole should lie in a common worldvolume direction of
 the $M5$-branes. One finds that to a single monopole one can add at most
 four $M5$-branes. An example of such a configuration is:
\be
\mbox{ {\scriptsize
$\left\{ \begin{array}{c|cccccccccc}
         \x & -  & \x & \x &  - & \x &  - & \x&  - & \x &  -  \\
         \x & \x &  - &  - & \x & \x &  - & \x&  - & \x &  -  \\
         \x & \x &  - & \x &  - & -  & \x & \x&  - & \x &  -  \\
         \x & \x &  - & \x &  - & \x &  - & - & \x & \x &  -  \\
         \x & \x & \x & \x & \x & \x & \x &A_7& A_8& z  &A_{10}            
                                         \end{array} \right.$} } 
\label{M5M5KK}
\ee
The harmonic functions depend only on the coordinate $x_{10}$.
 However, one may add more than one monopole to the four fivebranes.
 From (\ref{M5M5KK}) it is clear that the monopole could also
 have been placed with two components of the vector field in the 
 $(x_1, x_2),(x_3, x_4)$ or $(x_5, x_6)$ directions. In fact, in
 this way one can combine four monopoles with the four $M5$-branes:
\be
\mbox{ {\scriptsize
$\left\{ \begin{array}{c|cccccccccc}
         \x &  - & \x & \x &  - & \x &  - & \x&  - & \x &  -  \\
         \x & \x &  - &  - & \x & \x &  - & \x&  - & \x &  -  \\
         \x & \x &  - & \x &  - & -  & \x & \x&  - & \x &  -  \\
         \x & \x &  - & \x &  - & \x &  - & - & \x & \x &  -  \\
         \x & \x & \x & \x & \x & \x & \x &A_7& A_8&  z & A_{10}  \\  
         \x & \x & \x & \x & \x & B_5& B_6& \x& \x &  z & B_{10}  \\
         \x & \x & \x & C_3& C_4& \x & \x & \x& \x &  z & C_{10}  \\ 
         \x & D_1& D_2& \x & \x & \x & \x & \x& \x &  z & D_{10}            
                                         \end{array} \right.$} } 
\label{M5M5KKKK}
\ee
One may verify that this intersection is consistent
 with the $M5-\cal KK$ intersection rule 
 (\ref{M5KKwv}) and the $\cal KK - KK$ rule (\ref{KKKKa}). 

Having established the rule of how to add monopoles to an intersection
 of $M2$-branes and $M5$-branes or a mixture thereof, we are able to
 list all intersections involving $M2$-branes, $M5$-branes, waves and
 monopoles. It is enough to give only the intersection with the
 largest number of independent harmonics. All other intersections can
 be obtained from these by setting one or more of the harmonic
 functions  equal
 to one\footnote{This is not the case if one considers multiple
 monopoles of Type B and C.}.

The result is given in Table 4. The maximum number of intersecting
 objects $N$ equals  eight if we restrict ourselves to configurations 
 which can be reduced to pure $D$-brane intersections in $D=10$.
 We use the same notation as in \cite{mult}. In $D=11$ a configuration 
 is characterized by the number of $\x$'s (worldvolume directions) in
 each of the spatial coordinates. In this notation, the four
 fivebranes in (\ref{M5M5KK}) or 
 (\ref{M5M5KKKK}) are denoted by $[5^4]\{4,0,4,1\}$,
 since there are four coordinates with one $\x$, zero with two $\x$'s
 etc. In $D=10$ the same notation can be used, but then 
 a convention can be chosen to avoid giving $T$-dual solutions. The 
 convention is that in each coordinate $T$-duality should be used to 
 minimize the number of worldvolume directions. Then for $N=8$
 only four numbers need to be specified to characterize a $D=10$ class
 of (duality) equivalent solutions.
 In  Table 4 we have also indicated the unbroken 
 supersymmetry which directly follows from the unbroken supersymmetry
 of the corresponding $D$-brane intersection.

\vspace{3mm}
{\scriptsize
\begin{tabular}{|p{10mm}|p{40mm}|p{40mm}|p{40mm}|}
\hline
{\bf N=8} & {\bf(0,4,0,4)}${}_{SUSY=1/32}$ & {\bf(1,0,7,0)}${}_{SUSY=1/32}$ 
                       &{\bf(0,0,0,7)}${}_{SUSY=1/16}$   \\
\hline
&$[2^4,5^4]${\it\{0,4,0,5,0,0,0\}}
                &$[2^4,5^4]${\it\{1,0,6,1,1,0,0,0\}}
                          & $[2^3,5^4]${\it\{0,0,6,2,0,0,0\}}+$\cal{KK}$ \\
&$[2^3,5^4]${\it\{1,2,4,1,1,0,0\}}+$\cal{KK}$
                &$[2^3,5^4]${\it\{1,3,1,4,0,0,0\}}+ $\cal{KK}$
                          &$[2^1,5^4]${\it\{1,6,0,1,1\}}+3${\cal KK}$  \\
&$[2^2,5^4]${\it\{2,2,2,3,0,0\}}+2 $\cal{KK}$
                &$[2^2,5^4]${\it\{1,4,2,1,1,0,0\}}+2 $\cal{KK}$
                          &$[5^7]${\it\{0,0,0,7,0,0,1\}}+${\cal W}$  \\
&$[2^1,5^4]${\it\{0,4,2,2,0\}}+3 $\cal{KK}$
                &$[2^2,5^4]${\it\{0,2,4,2,0,0\}}+2 $\cal{KK}$
                          & \\
&$[5^4]${\it\{4,0,4,1\}}+4 $\cal{KK}$
                &$[2^1,5^4]${\it\{3,1,3,2,0\}}+3 $\cal{KK}$
                          & \\
&$[5^7]${\it\{0,3,0,4,0,1,1\}}+ $\cal{W}$
                &$[5^4]${\it\{0,6,0,2\}}+4 $\cal{KK}$
                          & \\
& 
                &$[5^7]${\it\{0,0,7,0,0,0,2\}}+ $\cal{W}$
                          & \\                   
&
                &$[5^7]${\it\{1,0,4,0,3,0,1\}}+ $\cal{W}$
                          & \\

\hline
\end{tabular}}

\vspace{3mm}
{\scriptsize
\noindent Table 4.\ {\bf N=8 intersections that reduce to pure
 $D$-brane intersections:} The boldface numbers indicate the ten
 dimensional T-duality class. The notation $[2^k,5^l] 
 + n{\cal KK}$ indicates that the intersections contain $k$
 $M2$-branes, $l$ $M5$-branes and $n$ monopoles. An additional
 wave is indicated by $+{\cal W}$.
}
\vspace{5mm}

Now consider using also the pair $(1|M2,{\cal W})$. The reduction to
 $D=10$ will then necessarily include also
 NS/NS branes\footnote{Such intersections were indicated by grey color
 in the Tables of \cite{mult}.}. 
 It turns out that there are three such maximum
 intersections. All other intersections follow by truncation of these ones.
 We find one intersection with $N=8$ and two intersections with
 $N=9$ independent harmonics:

\bea
N=8:&&[2^1,5^6]{\textstyle{\it \{1,0,4,3,0,0,1\}}}+{\cal W}\,, \nonumber\\
N=9:&&[2^1,5^7]{\textstyle{\it \{1,0,0,7,0,0,0,1\}}}+{\cal W}\,,\\
    &&[2^1,5^4]{\textstyle{\it \{1,6,0,1,1\}}} + 3 {\cal KK} + {\cal W} \,.
 \nonumber
\eea 

All three solutions have 1/32 unbroken supersymmetry. 
Interestingly enough we find intersections with {\it nine} independent
harmonics. These intersections have one common time direction, nine
relative transverse directions and one overall transverse direction.
They therefore naturally reduce, upon identifying all harmonics,
to a supersymmetric dilatonic 0-brane solution in two dimensions.
Since this solution involves the newly constructed $N=9$ intersection 
given above, it did not occur in our previous paper \cite{mult}.
The specific dilaton coupling in two dimensions is the same for each of the
two $N=9$ intersections since it only depends
on the number of independent harmonics (= field strenghts in two 
dimensions) \cite{Lu}. We find that the dilaton coupling
is given by $a=-4/9$.

The two intersections with $N=9$ are extensions of $N=8$ intersections
 with 1/16 supersymmetry in Table 4. The remaining
 intersection with 1/16 supersymmetry, $[2^3,5^4]+{\cal KK}$ 
 can also be
 extended to $N=9$, but this necessarily requires the use of a pair 
 from Table 2. For example, an additional fivebrane can be added,
 giving 1/32 supersymmetry.

\vspace{3mm}
\noindent{\bf 5. Conclusions}
\vspace{3mm}

In this letter we have considered intersections of $M2$-branes,
 $M5$-branes, waves and monopoles. We first considered the pair
 intersections, which fall in two groups (Table 1 and Table 2)
 depending on the coordinates on which the intersecting branes depend.
 Using only the pairs of Table 1, where the branes depend on overall 
 transverse coordinates,
 we then considered purely
 gravitational solutions with only monopoles and waves. We
 found three types of such intersections (see Table 3)
 consisting of multiple monopoles and in one case an additional
 wave. We next included the $M2$- and $M5$-branes and gave all
 intersections that can be reduced to ten-dimensional intersections
 involving only $D$-branes. This restriction is implemented by using only
 a limited number of the pair intersections of Table 1.
 This was completed by adding additional waves.
 As a new result we found two new configurations with {\it nine}
 independent harmonic functions. Upon reduction they  lead to a
 new supersymmetric 0-brane solution in two dimensions with
 dilaton coupling $a=-4/9$.

The pair intersections in Section 2 show the interesting feature that
 in some cases the reduction to $D=10$ gives rise to a solution
 which is not expressed in the standard way in terms of harmonic
 functions. In much of the previous work on pair intersections
 in $D=10$ the possibility of such solutions, which interpolate between
 standard harmonic single-brane solutions, but cannot themselves
 be  expressed in terms of two harmonic functions, was not considered
(see, however, \cite{Tse}). 
 These solutions may provide a useful hint in a search for more
 general, non-harmonic, pair intersections. In particular, it may well
 be that the structure of completely localized brane intersections
 can be clarified in this way.

In this letter we did not consider intersections containing
 multiple monopoles of Type B and C 
 where the $z$-isometry direction is not the same for all monopoles.
 Such configurations are characterized by the fact that, upon
 reduction to ten dimensions, they always lead to an intersection
 involving at least one monopole.
 Although the result can be derived in a straightforward manner
 it turns out that the answer is involved.
 This is due to the fact that for these cases not all possible
 configurations follow by truncation from the intersections
 with the maximum number of harmonics.

We finally note that we did not consider eleven dimensional
 intersections involving 9-branes. In order to do that, one
 should first be able to construct such a 9-brane solution. 
 We nevertheless found a hint in our calculations that the
 addition of such would-be 9-branes would be consistent with 
 supersymmetry\footnote{The use of supersymmetry in constructing
 multiple intersections is discussed in more detail in \cite{dR}.}
 in the following sense. Assuming that the
 unbroken supersymmetry of the 9-brane is determined by
\be
\bigl ( 1 + \gamma_{01\cdots 9}\bigr )\epsilon = 0\, .
\ee
\noindent we found that such a projection operator naturally follows by taking
 products of similar projection operators corresponding to the
 other eleven dimensional solutions.
 This suggests that to specific combinations of $M2$-, $M5$-branes,
 waves and monopoles a 9-brane can be added without breaking
 supersymmetry \cite{Gaunt}. It would
 be interesting to clarify the role of this would-be eleven dimensional
 9-brane.

\vspace{3mm}
\noindent{\bf Acknowledgements} 
\vspace{3mm}

This work is part of the research 
 program of the ``Stichting
 voor Fundamenteel Onderzoek der Materie'' (FOM). 
 It is also supported  by the European Commission TMR programme 
 ERBFMRX-CT96-0045,
 in which E.B. and M.~de R. are associated to the University of Utrecht.

\end{document}

%% file: M-A.pstex_t
\begin{picture}(0,0)%
\special{psfile=M-A.pstex}%
\end{picture}%
\setlength{\unitlength}{0.00083300in}%
\begingroup\makeatletter\ifx\SetFigFont\undefined%
\gdef\SetFigFont#1#2#3#4#5{%
  \reset@font\fontsize{#1}{#2pt}%
  \fontfamily{#3}\fontseries{#4}\fontshape{#5}%
  \selectfont}%
\fi\endgroup%
\begin{picture}(6388,1152)(322,-2377)
\put(1769,-2301){\makebox(0,0)[b]{\smash{\SetFigFont{11}{13.2}{\familydefault}{\mddefault}{\updefault}$D0$}}}
\put(2195,-2301){\makebox(0,0)[b]{\smash{\SetFigFont{11}{13.2}{\familydefault}{\mddefault}{\updefault}$W$}}}
\put(2622,-2301){\makebox(0,0)[b]{\smash{\SetFigFont{11}{13.2}{\familydefault}{\mddefault}{\updefault}$F1$}}}
\put(2989,-2301){\makebox(0,0)[b]{\smash{\SetFigFont{11}{13.2}{\familydefault}{\mddefault}{\updefault}$D2$}}}
\put(3416,-2301){\makebox(0,0)[b]{\smash{\SetFigFont{11}{13.2}{\familydefault}{\mddefault}{\updefault}$D4$}}}
\put(4206,-2301){\makebox(0,0)[b]{\smash{\SetFigFont{11}{13.2}{\familydefault}{\mddefault}{\updefault}$KK$}}}
\put(981,-2290){\makebox(0,0)[b]{\smash{\SetFigFont{11}{13.2}{\familydefault}{\mddefault}{\updefault}$IIA \ theory$}}}
\put(2195,-1425){\makebox(0,0)[b]{\smash{\SetFigFont{11}{13.2}{\familydefault}{\mddefault}{\updefault}${\cal W}$}}}
\put(3049,-1425){\makebox(0,0)[b]{\smash{\SetFigFont{11}{13.2}{\familydefault}{\mddefault}{\updefault}$M2$}}}
\put(3779,-1425){\makebox(0,0)[b]{\smash{\SetFigFont{11}{13.2}{\familydefault}{\mddefault}{\updefault}$M5$}}}
\put(4631,-1425){\makebox(0,0)[b]{\smash{\SetFigFont{11}{13.2}{\familydefault}{\mddefault}{\updefault}${\cal KK}$}}}
\put(980,-1425){\makebox(0,0)[b]{\smash{\SetFigFont{11}{13.2}{\familydefault}{\mddefault}{\updefault}$M \ theory$}}}
\put(4610,-2301){\makebox(0,0)[b]{\smash{\SetFigFont{11}{13.2}{\familydefault}{\mddefault}{\updefault}$D6$}}}
\put(3804,-2311){\makebox(0,0)[b]{\smash{\SetFigFont{11}{13.2}{\familydefault}{\mddefault}{\updefault}$S5$}}}
\put(6057,-2292){\makebox(0,0)[b]{\smash{\SetFigFont{11}{13.2}{\familydefault}{\mddefault}{\updefault}$flat\ space$}}}
\put(6057,-1424){\makebox(0,0)[b]{\smash{\SetFigFont{11}{13.2}{\familydefault}{\mddefault}{\updefault}$9\ brane$}}}
\put(5425,-2296){\makebox(0,0)[b]{\smash{\SetFigFont{11}{13.2}{\familydefault}{\mddefault}{\updefault}$D8$}}}
\end{picture}

%% file: final.bbl
\begin{thebibliography}{99}
%
\bibitem{Towns}P.~K.~Townsend, Phys. Let. {\bf B350} (1995) 184, 
               {\tt hep-th/9501068}.
%
\bibitem{Dabh}A.~Dabholkar, G.~W.~Gibbons, J.~A.~Harvey and F.~Ruiz-Ruiz, 
 Nucl.~Phys.~{\bf B340} (1990) 33.
%
\bibitem{Callan}C.~G.~Callan, J.~A.~Harvey and A.~Strominger,
 Nucl.~Phys.~{\bf B340} (1990) 611.
%
\bibitem{Duff-Lu}M.~J.~Duff and J.~X.~Lu,
 Nucl.~Phys.~{\bf B354} (1991) 141.
%
\bibitem{Duff-Stelle}M.~J.~Duff and K.~S.~Stelle, Phys. Lett. {\bf B253} 
(1991) 113.
%
\bibitem{Guven}R.~G\"uven, Phys. Lett. {\bf B276} (1992) 49.
%
\bibitem{Brink}H.~W.~Brinkmann, Proc.~Nat.~Acad.~Sci.~{\bf 9} (1923) 1.
%
\bibitem{Sorkin}R.~D.~Sorkin, Phys.~Rev.~Lett. {\bf 51} (1983) 87;
 D.~J.~Gross and M.~J.~Perry, Nucl.~Phys.~{\bf B226} (1983) 29.
%
\bibitem{Pol2}J.~Polchinski and E.~Witten,
 Nucl.~Phys.~{\bf B460} (1996) 525,
 {\tt hep-th/9510169}.
%
\bibitem{Be2}E.~Bergshoeff, M.~de Roo, M.~B.~Green, G.~Papadopoulos and
 P.~K.~Townsend,
 Nucl.~Phys.~{\bf B470} (1996) 113,
 {\tt hep-th/9601150}.
%
\bibitem{Howe}P.~S.~Howe and E.~Sezgin, 
Phys.~Lett.~{\bf B390} (1997) 133,  {\tt hep-th/9607227}.
\bibitem{Papa2}G.~Papadopoulos and P.~K.~Townsend, 
Phys.~Lett.~{\bf B393} (1997) 59,
 {\tt hep-th/9609095}.
\bibitem{Pol1}J.~Polchinski, {\sl TASI-lectures on $D$-branes}, 
 {\tt hep-th/9611050}.
\bibitem{Duff}M.~J.~Duff, {\sl Supermembranes}, TASI lectures, Boulder 1996,
 {\tt hep-th/9611203}.
%
\bibitem{mult}E.~Bergshoeff, M.~de Roo, E.~Eyras, B.~Janssen and J.~P.~van
 der Schaar, 
 {\it Multiple intersections of D-branes and M-branes}, 
 {\tt hep-th/9612095}, to appear in Nucl.~Phys.~B.
%
\bibitem{Papad-Towns1}G.~Papadopoulos and P.~K.~Townsend, 
  Phys.~Lett.~{\bf B380} (1996) 273,
 {\tt hep-th/9604068}.
%
\bibitem{Tseyt1}A.~A.~Tseytlin, 
 Nucl.~Phys.~{\bf B475} (1996) 149,
 {\tt hep-th/9604035}.
%
\bibitem{BeBeJa}K.~Behrndt, E.~Bergshoeff, B.~Janssen, 
 Phys.~Rev.~{\bf D55} (1997) 3785, 
 {\tt hep-th/9604168}.
%
\bibitem{Gaunt}J.~Gauntlett, D.~Kastor and J.~Traschen, 
 Nucl.~Phys.~{\bf B478} (1996) 544, 
 {\tt hep-th/9604179}.
%
\bibitem{Tseyt2}A.~A.~Tseytlin, 
 Nucl.~Phys.~{\bf B487} (1997) 141, {\tt hep-th/9609212}.
%
\bibitem{Are}I.~Ya.~Aref'eva and A.~Volovich,
 {\it Composite p-branes in Diverse Dimensions},
 {\tt hep-th/9611026}.
%
\bibitem{Khvien}N.~Khviengia, Z.~Khviengia, H.~L\"u and C.~N.~Pope,
 Phys.~Lett.~{\bf B388} (1996) 313,
 {\tt hep-th/9605082}.
%
\bibitem{Costa}M.~Costa, {\it Composite M-branes}, {\tt hep-th/9609181},
to appear in Nucl.~Phys.~B.
%
\bibitem{Tse} A.A.~Tseytlin, {\it Composite BPS configurations of
p-branes in 10 and 11 dimensions}, {\tt hep-th/9702163}.
%
\bibitem{Tnew} J.P.~Gauntlett, G.W.~Gibbons, G.~Papadopoulos and
 P.K.~Townsend, 
 {\it Hyper-K\"ahler Manifolds and Multiply Intersecting Branes}, 
 {\tt hep-th/9702202}.
%
\bibitem{Tseyt3}J.G.~Russo and A.A.~Tseytlin, 
 {\it Waves, boosted branes and BPS states in M-theory}, 
 {\tt hep-th/9611047}.
%
\bibitem{Lu}H. L{\" u}, C.~N.~Pope, E.~Sezgin and K.~S.~Stelle, 
 Phys.~Lett.~{\bf B371} (1996) 46, 
 {\tt hep-th/9511203}; 
 Nucl.~Phys.~{\bf B456} (1995) 669, 
 {\tt hep-th/9508042}.
%
\bibitem{dR}M.~de Roo, 
 {\it Intersecting branes and supersymmetry},
 presented at 
 {\it Supersymmetry and Quantum Field Theory},  
 International Seminar dedicated to the memory of D.~V.~Volkov,
 Kharkov, Ukraine (1997), 
 {\tt hep-th/9703124}.

\end{thebibliography}
